\documentclass{pasj01}

\Received{$\langle$reception date$\rangle$}
\Accepted{$\langle$acception date$\rangle$}
\Published{$\langle$publication date$\rangle$}
\newcommand{\hii}{H~II~}

\usepackage{soul}
\usepackage{color}

\begin{document} 

\Received{}
\Accepted{}
\Published{}
\title{Proper Motions of Water Masers in the Star-forming Region IRAS 23139+5939\thanks{Based on observations conducted with the KaVA; the combination array of the Korean VLBI Network (KVN) and Japanese VLBI Exploration of Radio Astrometry (VERA), via a KASI and NAOJ agreement). https://radio.kasi.re.kr/kava/main\_kava.php}
\thanks{As section of the thesis to be submitted by Toledano--Ju\'arez as a partial fulfillment for the requirements of Ph. D. Degree in Physics, Doctorado en Ciencias (F\'isica), CUCEI, Universidad de Guadalajara}}

\author{Miguel A. \textsc{Trinidad}\altaffilmark{1}%
\thanks{Corresponding authors: eduardo.delafuente@academicos.udg.mx, hiroimai@km.kagoshima-u.ac.jp, and trinidad@astro.ugto.mx}}
\altaffiltext{1}{Departamento de Astronom\'{i}a, Universidad de Guanajuato, Apartado Postal 144, 36000, Guanajuato, Guanajuato, M\'exico}
\email{trinidad@astro.ugto.mx}

\author{Hiroshi \textsc{Imai},\altaffilmark{2,3,$\ddagger$}}
\altaffiltext{2}{Amanogawa Galaxy Astronomy Research Center, Graduate School of Science and Engineering, Kagoshima University, 
1-21-35 Korimoto, Kagoshima 890-0065, Japan}
\altaffiltext{3}{Center for General Education, Institute for Comprehensive Education, Kagoshima University,  ¥¥ 1-21-30 Korimoto, Kagoshima 890-0065, Japan}
\email{hiroimai@km.kagoshima-u.ac.jp}

\author{Eduardo \textsc{de la Fuente}\altaffilmark{4,5,6$\ddagger$}}
\altaffiltext{4}{Departamento de F\'{i}sica, CUCEI, Universidad de Guadalajara, Blvd. Marcelino Garc\'{i}a Barragan 1420, Ol\'{i}mpica, 44430, Guadalajara, Jalisco, M\'exico}
\email{eduardo.delafuente@academicos.udg.mx}

\altaffiltext{5}{Institute for Cosmic Ray Research (ICRR), University of Tokyo, 1--5 Kashiwanoha 5-Chome, Kashiwa, Chiba 277-8582, Japan}

\altaffiltext{6}{Maestr\'ia en Ciencias de Datos, CUCEA, Universidad de Guadalajara, Perif\'erico Norte 799 N\'ucleo Universitario, Los Belenes, 45100 Zapopan, Jalisco, M\'exico}

\author{Ivan \textsc{Toledano-Ju\'arez}\altaffilmark{6,7}}
\altaffiltext{7}{Doctorado en Ciencias F\'{i}sicas, CUCEI, Universidad de Guadalajara, Blvd. Marcelino Garc\'{i}a Barragan 1420, Ol\'{i}mpica, 44430, Guadalajara, Jalisco, M\'exico}
\email{toledano.ivan16@gmail.com}

\author{Joseph M. \textsc{Masqu\'e}\altaffilmark{1}}
\email{jmasque@ugto.mx}

\author{Tatianna \textsc{Rodr\'iguez-Esnard}\altaffilmark{8}}
\altaffiltext{8}{Instituto  de  Geof\'isica y Astronom\'ia de la Republica de Cuba, La Habana, Cuba.}
\email{tatiana@iga.cu}

\KeyWords{ISM: jets and outflows --- stars: protostars --- masers --- ISM: individual objects (IRAS 23139+5939) --- methods: data analysis}

\maketitle

\begin{abstract}

We observed H$_2$O (6$_{16}$ $\rightarrow$ 5$_{23}$) maser emission associated with the high-mass star-forming region IRAS 23139+5959 using the KaVA a combination of VLBI arrays between the KVN (Korea) and VERA (Japan). Through multi-epoch KaVA observations, we detected three groups of maser features, two of which coincide with those previously detected by the Karl G. Jansky Very Large Array (VLA). By determining the maser proper motions, we found that the first of maser groups exhibits an expanding motion that traces a wide-angle outflow almost along the line of sight, while the second one seems to be associated with the envelope of an \hii region. We discuss the star formation activity in IRAS 23139+5939, which may be reflected in the high variability of H$_2$O masers associated with an outflow seen from the front.

\end{abstract}

\section{Introduction}

Massive young stellar objects (YSOs) are deeply immersed in a large amount of circumstellar material. Consequently, these cannot be observed directly at optical or even in some cases at near-infrared wavelengths. Fortunately, most embedded YSOs can be detected with radio telescopes at centimeter wavelengths, and this allows us to investigate their nature and study their formation process. However, due to their relatively low flux density in radio continuum emission, it has been difficult to perform a detailed study of the continuum emission except in limited cases. For example, this emission is mainly associated with the signposts of newly born massive stars, such as radio jets and \hii regions. In addition to this, there is no entirely accepted model to explain the formation process of massive stars. Because of this reason, additional observational techniques are required, such as observations with a high angular resolution, allowing us to study the gas very close to massive YSOs. 

Within this context, observations of maser transitions of several molecular species, particularly water (H$_2$O), in the proximity to massive YSOs provide a powerful diagnostic tool to investigate the first stages of massive star formation. H$_2$O maser emission is a selective tracer that can occur either in molecular outflows (e.g., \cite{Chernin1995,Torrelles1998,Furuya2000,Patel2000,Torrelles2014}) or, in rare cases, such as circumstellar disks (e.g., \cite{Shepherd1999,Seth2002,Rodriguez-Esnard2014}). Maser features (a cluster of spots in individual velocity channels) are like  ``test particles'', tracing the masing gas' three-dimensional velocity field very close to YSOs. The combination between line-of-sight velocities and the proper motions of maser features is crucial for determining whether a group of features is rotating, contracting, or expanding, without any assumption of the geometry. The frequent tracking of the masers in temporal scales from few weeks to months is necessary to minimize the time evolutionary effects of the maser features themselves and to be able to calculate their reliable proper motions (e.g., \cite{Chibueze2012}). 

Maser motions may be useful to study the stability of the continuum jet and the possible entrance of material at the interface between the jet and the surrounding interstellar material. One also expects to see maser features moving along the outflows, tracing circumstellar disks, or delineating the edge of isotropic gas ejections, etc. The new detection of an outflow activity may allow us to identify a new and previously unseen center of massive star formation activity (e.g., \cite{Torrelles2001,Torrelles2002}). In the earliest stages of massive YSO's evolution, some of these centers are likely associated with unexpected phenomena such as ``short-lived" episodic ejection events characterized by a few decades' kinematic age.  The powering mechanism of non-collimated outflows associated with such episodic events is not still clear in current star formation theories. These facts open new challenges in the study of the earliest stages of stellar evolution (e.g., \cite{Torrelles2011, Trinidad2013}). In order to see the fates of the short-lived events, we should monitor the isotropic gas ejections for several years.

IRAS~23139+5939 (G111.25$-$0.77) is a high-mass star-forming region with a luminosity of $\sim 2 \times 10^4\; L_{\odot}$ \citep{Sridharan2002}{\footnote{Following \citet{Lang1992}, an estimated mass from this bolometric luminosity of $\sim$ 18 M$_\odot$ was calculated by \citet{Trinidad2006}. However, due to IRAS fluxes were used to derive this estimation, considerable uncertainties must be taken into account.}}. It is located near the S157 region at a distance of 3.4~kpc \citep{Choi2014}. This region has been studied in radio continuum  and with different molecular tracers \citep{Tofani1995,Trinidad2006,Wu2010,Obonyo2019}. Methanol maser studies have been performed by several authors \citep{Szymczak2018,Durjasz2019} showing high variability on time. On the other hand, IR and sub-mm observations are reported by UKIRT \citep{Varricatt2010}, WISE \citep{Wright2010}, MSX \citep{Urquhart2011}, BGPS \citep{Ellsworth2015}, and SCUBA \citep{DiFrancesco2008} confirming its nature as YSO. CO observations reveal intense emission by tracing molecular outflow activity (\cite{Wouterloot1989,Beuther2002}). Using tracers of molecular outflow emission such as SiO~($J=2\rightarrow 1$), SiO~($5\rightarrow 4$), and HCO$^{+}$~($1\rightarrow 0$), \citet{Sanchez2013} found extended high-velocity wings. However, a bipolar morphology has not been revealed in detail. The combination of new results in this paper with those of previous observations of H$_2$O masers \citep{Tofani1995,Trinidad2003,Goddi2005,Trinidad2006,Choi2014} allows us to clarify such an outflow morphology.

This region has also been studied by \citet{Trinidad2006} using the VLA in A-configuration at 1.3 and 3.5 cm. They found two radio continuum sources (I23139 and I23139S) at wavelengths of 3.5 cm and two H$_2$O maser groups. The latter is spatially associated with the continuum sources, obtaining an accuracy of $\sim$10 mas in relative positions between I23159 and the maser features.  The most interesting result comes from the spatial distribution of the water maser group associated with the strongest continuum source (I23159); the maser  distribution exhibits a shell-like structure, very close to the continuum source's peak emission. These masers have blue-shifted velocities and are not gravitationally bound. Therefore, these could trace expansion or contraction. The expansion scenario seems to be the most favorable because the continuum source, apparently hosting maser features, could be a thermal jet or stellar wind. Furthermore, I23159 also seems to be the center of a CO molecular outflow observed in the region. 

Besides, other maser group is spread out on a strip of $\sim$0.2\arcsec\ and seems to be aligned in the northeast--southwest direction pointing toward the weaker continuum source (I23139S). This group is composed of three maser features, two of them have blue-shifted velocities, and the other one has a velocity close to the source's systemic velocity ($-44.1$~km~s$^{-1}$, \cite{Bronfman1996}). The spatial distribution of these features also could trace an outflow.

In order to build the three-dimensional velocity field of the H$_2$O maser features in IRAS 23139+5939 and investigate their nature, in this paper we present VLBI multi-epoch observations carried out with KaVA, the combined VLBI array of three 21~m telescopes of the KVN (Korean VLBI Network; \cite{Lee2014}) and four 20~m telescopes of VERA (VLBI Exploration of Radio Astrometry; \cite{Kobayashi2003}). We detected three maser groups, two of them associated with those detected with the VLA: Groups 1 and 2). Based on their calculated proper motions, we found that the water maser features spatially associated with the continuum source I23159 (Group 1) are tracing expansion motions, while the maser features located to the southeast of I23139 (Group 2) seem to be associated with a H~II region.

In Section \S~\ref{sec:obs}, we describe the KaVA observations for maser proper motion measurement, while results and discussion are shown in sections \S~\ref{sec:res} and \S~\ref{sec:disc}, respectively. Main conclusions are given in \S~\ref{sec:summary}.

\section{Observations and data analysis}
\label{sec:obs}

We observed H$_2$O masers in IRAS 23139+5939 with seven telescopes of KaVA (proposal ID KaVA18A-05) in three sessions in 2018; on February 6, March 29, and May 22, each one with a duration of 6 hr. As the seven KaVA telescopes form baselines from 300 to 2300~km, KaVA is relatively sensitive to extended radio emission. This has been demonstrated in the earliest publications (e.g.,\cite{Niinuma2014,Matsumoto2014}). The unique capabilities of high-accuracy astrometry with dual beams and four frequency-band simultaneous observations have been also highlighted in VERA (\cite{Hirota2020}) and KVN (e.g., \cite{Yoon2018}), respectively, and these are nowadays applicable to these KaVA observations. However, their applications are out of this paper based on observations in the KaVA's early phase.  

In order to perform the KaVA observations, continuum sources, 3C454.3, J2312+7241, and J0019+7327 were used as flux density, instrumental delay/phase and band-pass calibrators, respectively. Parameters of the KaVA observations are given in Tab. \ref{tab:tab1}. Received signals were digitized at a rate of 2 Gbps, filtered into 16 base-band channels (BBCs), each one with a bandwidth of 16 MHz, and recorded at a rate of 1 Gbps. One of the BBCs covered a velocity width of 210~km~s$^{-1}$, including an LSR velocity range from $\sim$--65 to 0 km~s$^{-1}$ to detect water masers. Recorded signals were processed using the Daejeon Correlator in the Korea-Japan Correlation Center to yield 2048 spectral channels per BBC, corresponding to a velocity spacing of $\sim$ 0.11~km~s$^{-1}$. 

We reduced the data using the standard procedures of the NRAO AIPS software package, which was handled by the Python/ParselTongue\footnote {See: https://www.jive.eu/jivewiki/doku.php?id=parseltongue:parseltongue .} scripts dedicated to the KaVA data. We calibrated the visibility amplitudes using the corresponding antenna gains and system temperature tables.

Instrumental clock parameters and complex band-pass characteristics were determined and calibrated using the scans in the three calibrators mentioned above. We adopted a rest frequency of the H$_2$O 6$_{16}$ $\rightarrow$ 5$_{23}$ maser transition line to be 22.23508 GHz. The data were cross-calibrated in phase and amplitude using the velocity channel including the most compact and the brightest maser spots at V$_{\rm LSR}=-$47.0, $-$56.7, and $-$56.7~km~s$^{-1}$ in the three sessions, respectively, as reference. For sessions 2 and 3, the maser spot used as reference was the same, but it was not detected for session 1, so another was used instead (the most intense at that time), located about 210~mas further away from that of sessions 2 and 3. Calibrated visibilities were naturally weighted to produce the synthesized beam of 2.1 $\times$ 1.4 [mas] at a position angle of $-26.6^{\circ}$ (for session 2).

These procedures allowed us to generate the synthesis maps of the maser emission with a root-mean-square noise ranged from $\sim$ 50~mJy to $\sim$ 350~mJy, depending on the session and the existence of bright maser spots in the map. The dynamic range of the image cube (the highest ratio of the peak intensity of a maser spot to the noise level in the same velocity channel) achieved 365, while maser spots with a signal-to-noise ratio higher than 8 were identified. In this condition, we obtained a relatively positional accuracy of about 0.1 mas, which will give us a reliable estimation of maser proper motions.

Maser spots brighter than 0.4~Jy were found automatically within an area of 0.4\arcsec$\times$0.4\arcsec.  All identified maser spots were once fitted with two dimensional elliptical Gaussians, by determining their positions, flux densities, and radial velocities. Then we further flagged out carefully by eye false maser spots raised by bright side lobes around bright maser spots. The remained spots  were grouped into maser features; each one of them corresponds to a physical gas clump composed of maser spots located within $\lesssim$1 mas in the sky and $\lesssim$2 km~s$^{-1}$ in velocity. We only chose maser features that host at least four maser spots and that are isolated in the sky and velocity. Nevertheless, it was difficult to flag out all possible false features and identify all reliable features, in which some fainter but real features close to brighter features were flagged out while some isolated but false features could contaminate.

We determined the absolute coordinates of the maser spots used as reference mentioned above using the fringe-rate mapping method \citep{Walker1981}. For the reference spot observed in the second session, when it was the brightest in the three sessions, we obtained its absolute coordinates: R.A.(J2000)$=$23$^{\rm h}$16$^{\rm m}$10$^{s}$\hspace{-2pt}.3676$\pm 0^{\rm s}$\hspace{-2pt}.016, Dec (J2000)$=+$59$^{\circ}$55$^{\prime}$28$^{\prime\prime}$\hspace{-2pt}.43$\pm 0^{\prime\prime}$\hspace{-2pt}.10. The coordinate uncertainty was limited to $\sim$0.1\arcsec\ due to the relatively low flux densities of reference spots, and sparse fringe rate calibration using scans on calibrators observed every $\sim$30 min. Nevertheless, this is enough to associate them with the radio continuum source detected in the region. All maser feature position offsets were measured relative to the stronger maser spot detected in the second session of KaVA observations.

Since the strongest maser feature, which was used as reference for the calibration, was not persistent in all three sessions (it was not detected in session 1 and other maser feature was selected), the strongest one was not used as positional reference to align the maps and other persistent features were identified. In order to align the maps, two maser features {\bf of Group 1}, persistent in all three sessions, were selected, whose radial velocity and position (for session 2) are about $-$42.6 and $-$35.2~km s$^{-1}$, and (17.65, 116.04) and  (–5.92, 3.81)~mas, respectively. The reference position to align the maps was determined by calculating the mean position (R.A. and Dec) of both selected maser features, being ($5.87\pm0.01$, $59.93\pm0.02$)~mas (R.A.(J2000)$=$23$^{\rm h}$16$^{\rm m}$10$^{s}$\hspace{-2pt}.3684$\pm 0^{\rm s}$\hspace{-2pt}.0008, Dec (J2000)$=+$59$^{\circ}$55$^{\prime}$28$^{\prime\prime}$\hspace{-2pt}.492$\pm 0^{\prime\prime}$\hspace{-2pt}.060; see Tab. \ref{tab:tab2}).

Given that the calibration of the second and third sessions was performed using the same maser feature, we first aligned the maps of these two sessions, and then the map of the first session was adjusted. In order to do this, we aligned the mean position of the two maser features selected in the third session, assuming that these have been moved at a constant velocity from positions in the second session, i.e. for 54 days (time separation between the observations of sessions 2 and 3). Finally, assuming that these features between sessions 2 and 3 are moved with constant velocity, this velocity was calculated using the mean position and the time between both sessions. Accordingly, considering a time span of 51 days between the first and second session, the estimated velocity between sessions 2 and 3 was also applied to the mean position of the two maser features selected in the first session. Thus, after aligning the maps, all the positions of the maser features from the three observed sessions were determined. Then, some of the features detected in two and/or three epochs were selected to calculate their proper motion.

\begin{longtable}{cllccc}
  \caption{Observational parameters}
  \label{tab:tab1}
  \hline              
  Source & Right Ascension & Declination & Flux density & Observing time & Source \\
  name & (J2000) & (J2000) & (Log in Jy) & (s) & Type \\ 
\endfirsthead
\hline
\endhead
  \hline
\endfoot
  \hline
\endlastfoot
  \hline
  3C454.3 & 22$^{\rm h}$53$^{\rm m}$57$^{\rm s}$\hspace{-2pt}.747937 & +16$^{\circ}$08\arcmin 53\arcsec\hspace{-2pt}.56094 & 9.63 & 150$\times$2 & Fringe finder \\
  J2312197$+$724127 & 23$^{\rm h}$12$^{\rm m}$19$^{\rm s}$\hspace{-2pt}.697840 & +72$^{\circ}$41\arcmin 26\arcsec\hspace{-2pt}.91754 & 0.40 & 180$\times$12 & Calibrator \\
  IRAS~23139$+$5939 & 23$^{\rm h}$16$^{\rm m}$10$^{\rm s}$\hspace{-2pt}.337 & +59$^{\circ}$55\arcmin 28\arcsec\hspace{-2pt}.61  & 30,000 & 1500$\times$12 & Target \\
  J0019458+732730 & 00$^{\rm h}$19$^{\rm m}$45$^{\rm s}$\hspace{-2pt}.786387 & +73$^{\circ}$27\arcmin 30\arcsec\hspace{-2pt}.01750 & 0.89 & 150$\times$1 & Fringe finder \\ 
\end{longtable}

\section{Results}
\label{sec:res}

\begin{figure*}
 \begin{center}
    \includegraphics[width=\textwidth]{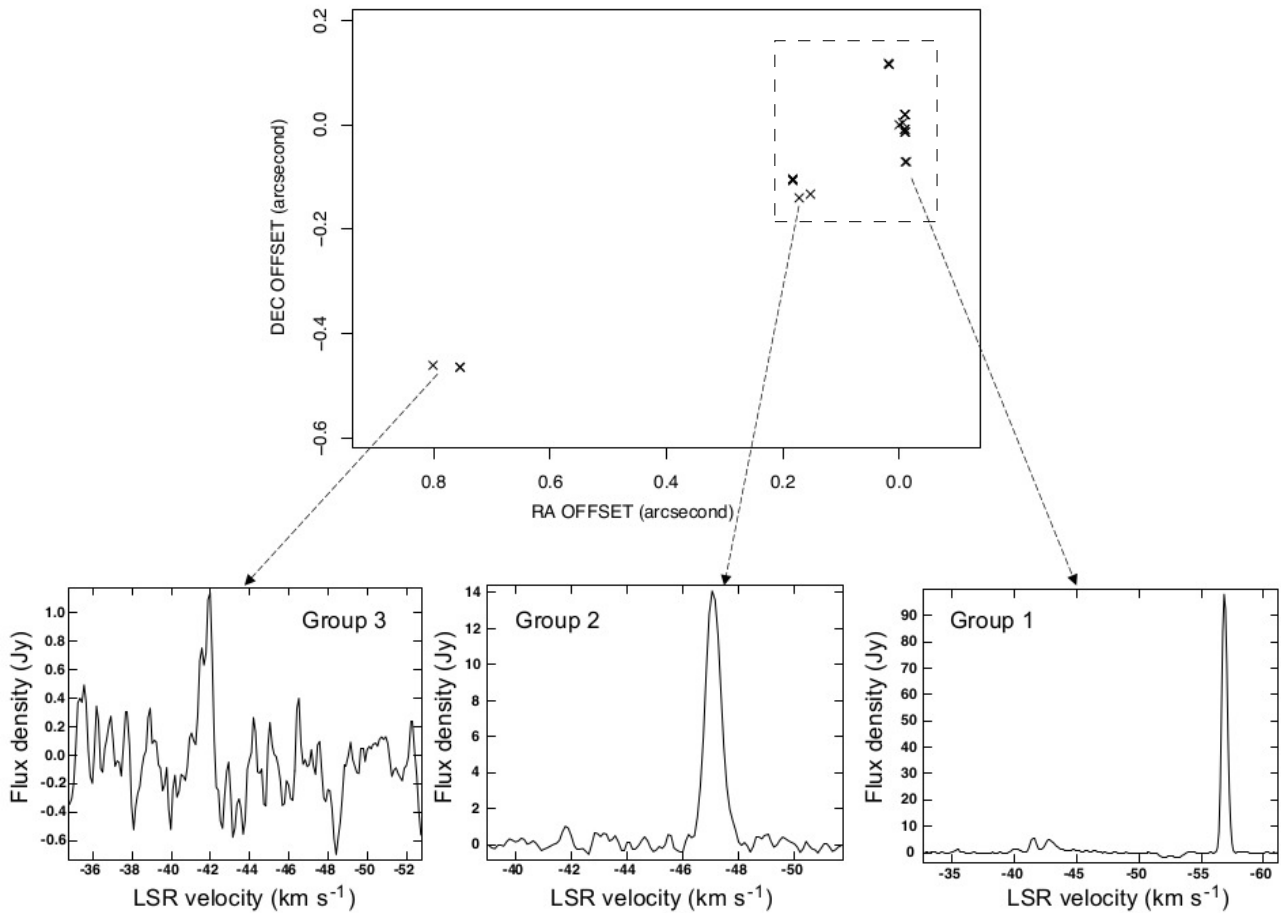}
 \end{center}
 \caption{{\it Top}: Distribution of H$_2$O maser features detected with KaVA during session 2. Maser distribution is centered in the strongest maser feature detected in the field at V$_{\rm LSR}=-$56.7~km~s$^{-1}$. The dashed square, which is showed as inset, refers to Fig. \ref{fig:ProperMotions}a) field of view. {\it Bottom}: Spectra of the three groups of maser features detected during session 2.}
 \label{fig:Epoch2}
\end{figure*}

We identified 19, 23, and 13 H$_2$O maser features toward IRAS~23139$+$5939 in the three observing sessions, respectively, whose velocity range spans from V$_{\rm LSR}$ = $-$57 to $-$33~km s$^{-1}$. Table \ref{tab:tab2} gives the parameters of the identified H$_2$O maser features  and Fig. \ref{fig:Epoch2} shows the distribution of all the maser features detected during session 2 toward IRAS~23139$+$5939. We notice that the maser features are distributed in three groups, labeled as 1, 2, and 3. Group 1 contains the most of maser features, while the furthest one, Group 3 has few maser features. This behavior was observed practically in the three sessions. Specifically, H$_2$O maser features in Groups 1 and 2 were detected in all sessions, whereas those of Group 3 were only detected in the first and second sessions.

In general, the peak intensity of the maser features range from a few tenths to a few tens of Jy beam$^{-1}$. However, a powerful maser feature was detected in the second and third session, with a peak intensity of $\sim$ 100 and $\sim$ 67 Jy beam$^{-1}$, respectively (Tab. \ref {tab:tab2}). This maser feature, with a velocity of $-$56.8~km~s$^{-1}$, is blueshifted with respect to the systemic velocity ($-$44.1~km~s$^{-1}$) of the IRAS~23139+5939 region and was used as reference maser for calibration, but it was not detected in the first session. Fig. \ref{fig:Epoch2} also shows the spectra of the three maser groups from the second session, in which the maser emission was the most populated. Most of these maser features are blueshifted and the strongest ones, except in Group 3. A similar trend is also observed in other sessions. This trend is consistent with the variability of the H$_2$O maser emission toward IRAS~23139$+$5939 studied by \citet{Trinidad2003}.

\begin{figure}
\includegraphics[width=0.5\columnwidth]{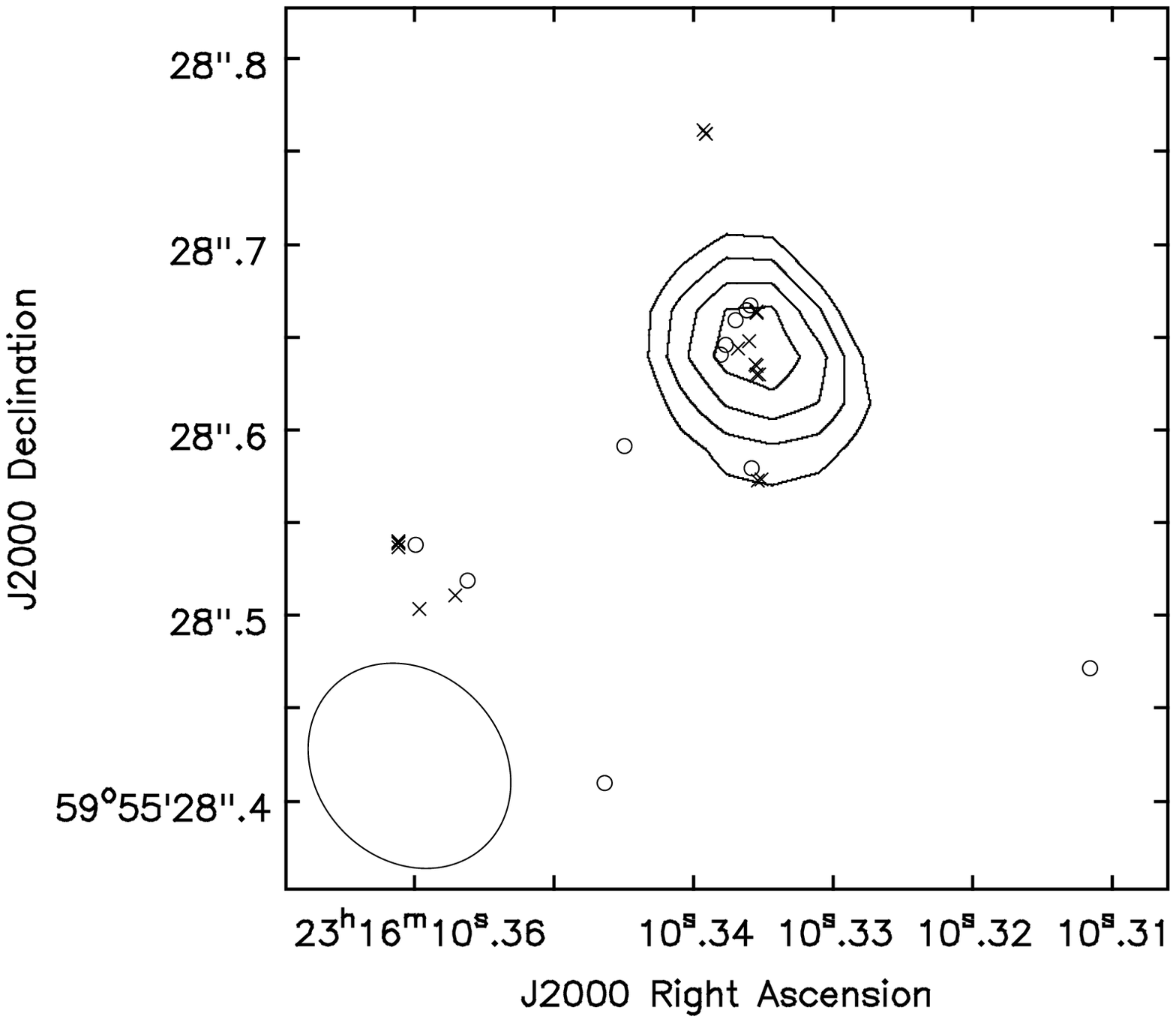}
\includegraphics[width=0.5\columnwidth]{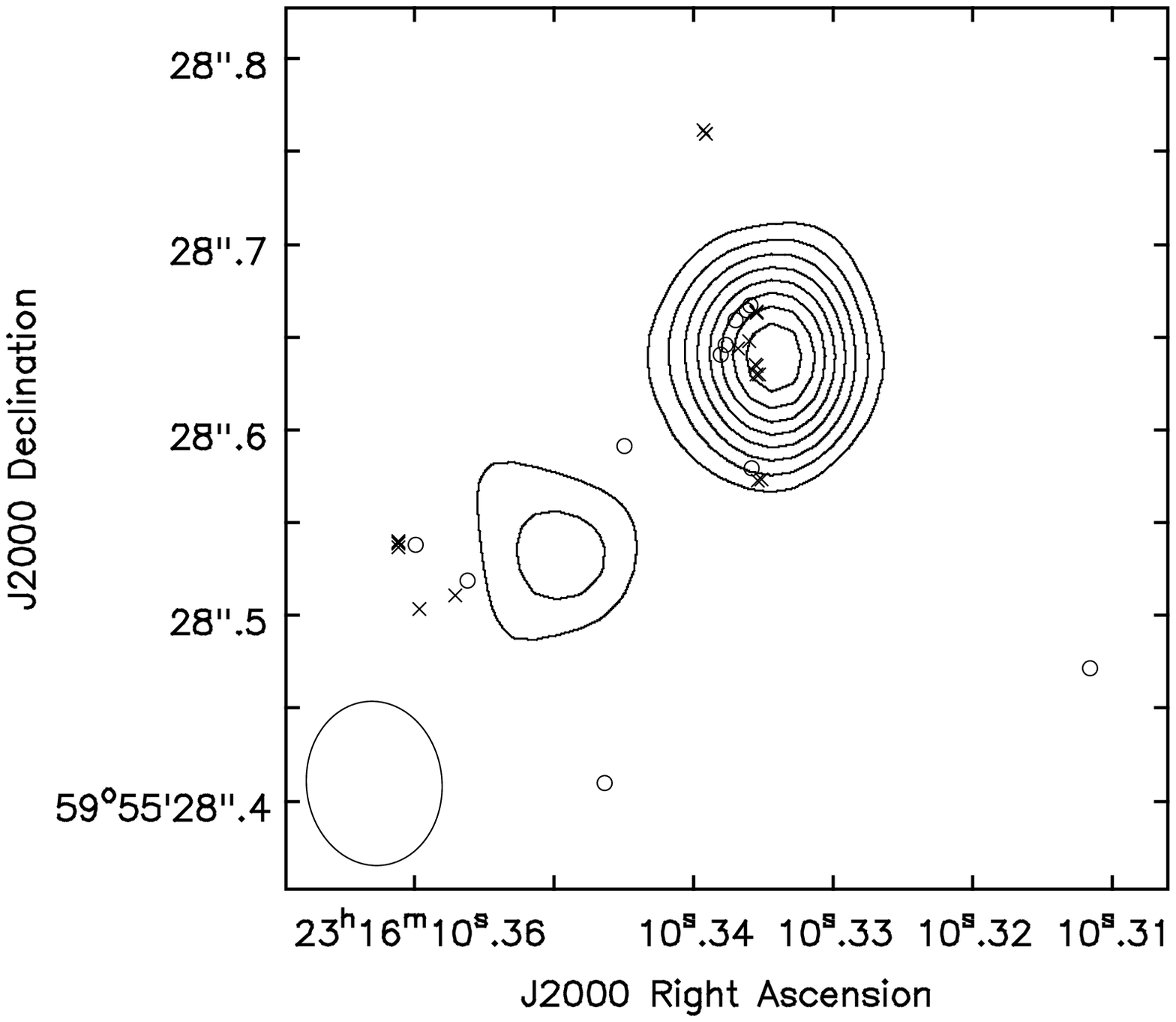}
\caption{Left: Distribution of H$_2$O  maser features around the 1.3~cm continuum source I23139 (contours) mapped by  \citet{Trinidad2006}. Contours are -4,-3,3,4,5,6 $\times$ 0.14 mJy beam$^{-1}$ (the rms noise). The beam size of 0.12''$\times$0.10'' with PA of 42.69$^{\circ}$ is shown at the bottom left. Open circles denote maser features detected by \citet{Trinidad2006} with the VLA, while cross symbols denote maser features found in our KaVA observations (groups 1 and 2 of Session 2), whose coordinates are shifted by ($-$0\arcsec.23, 0\arcsec.21) from those determined in the fringe-rate mapping method (see Section \ref{sec:obs}) for comparison with the VLA maser distribution. Right: Same as the left image but for observations reported by \citet{Moscadelli16}. Contours are -0.2,0.2,0.3,0.4,0.5,0.6,0.7,0.8,0.9 $\times$ 0.20 microJy beam$^{-1}$ peak flux, with a rms of 10 $\mu$Jy beam$^{-1}$. In order to match with the peak emission of the continuum source detected by \citet{Trinidad2006}, a shift of (0\arcsec.027, 0\arcsec.013) was applied. The beam size of 0.09''$\times$0.07'' with PA of 5.23$^{\circ}$ is shown at the bottom left. At the SE, the second source discussed in the text is clearly observed. The group of masers at the extreme left (group 2) could be related to this continuum source.}
 \label{fig:Epoch2-VLA}
\end{figure}

Fig. \ref{fig:Epoch2-VLA} shows the H$_2$O maser features superimposed in the contour map of the 1.3~cm continuum emission from I23139 detected with the VLA \citep{Trinidad2006}, as well as the maser features in Groups 1 and 2 detected in the second session of the KaVA observations. The absolute coordinates of these maser features detected with KaVA do not match exactly with those found with the VLA, which could be explained by their uncertainty (actually worse than $\sim$0.1$^{\prime\prime}$, see Section \ref{sec:obs}). 

In order to carry out a visual comparison between the maser feature distributions in the KaVA observations with those in the VLA, a shift of ($-$0\arcsec.23, 0\arcsec.21) was applied to the position of the KaVA maser features determined with the fringe-rate mapping method (see Fig. \ref{fig:Epoch2-VLA}). Comparing both maser distributions, we found that the maser features of the central region of Group 1 (of the KaVA observations) seem to be related to the cluster of the masers (detected with the VLA)  spatially associated with the radio continuum source I23139. In addition, we found that the maser features of Group 2 are associated with those H$_2$O masers found in the southeast from I23139 with the VLA. However, as it was mentioned above, this comparison is only for visual purposes, and because the offset applied to the KaVA data is more significant than its positional uncertainty, the findings from this comparison should be carefully considered. 

On the other hand, the maser features in Group 3 were not detected in the previous VLA observations or in the third session of the KaVA observations. We could explain the latter by the fact that H$_2$O  masers observed in star-forming regions show significant change in their global distribution in long and even short time scales (e.g., \cite{Trinidad2003}). 

We calculated the relative proper motions of the maser features detected in two and/or three sessions by assuming that these have linear motions and by applying a linear fitting to their positions. 
Since the H$_2$O maser emission turned out to be highly variable on a time scale of few weeks, it was only possible to measure a limited number of maser proper motions, nine in total in all IRAS~23139+5939 region (seven in Group 1 and two in Group 2) (see Tab. \ref{tab:tab3}). 
The determined geometric center (``center of motion'') of the maser features having proper motion in Group 1 is R.A.(J2000)$=$23$^{\rm h}$16$^{\rm m}$10$^{s}$\hspace{-2pt}.3676$\pm 0^{\rm s}$\hspace{-2pt}.0003, Dec (J2000)$=+$59$^{\circ}$55$^{\prime}$28$^{\prime\prime}$\hspace{-2pt}.432$\pm 0^{\prime\prime}$\hspace{-2pt}.042, corresponding to a proper motion of (0, 0)~mas~yr$^{-1}$. 
Calculated proper motions seem to show that all maser features are being moved northward. However, this result is not consistent with previous studies (e.g., \cite{Choi2014}). The mean of the north-south components of the proper motions is $\sim0.7$~mas~yr$^{-1}$ (1~mas~yr$^{-1}$ = 16~km~s$^{-1}$), which is larger than that reported by other authors (e.g. $-2.3$~mas~yr$^{-1}$: \cite{Choi2014}). We suppose that this trend is not real and could be attributed to the reference frame moving together with the reference feature. In order to remove this bias, the mean of the north-south components {\bf $\sim$ 0.7~mas~yr$^{-1}$} was subtracted from all maser proper motions.

\begin{figure*}
 \begin{center}
\includegraphics[width=0.845\textwidth]{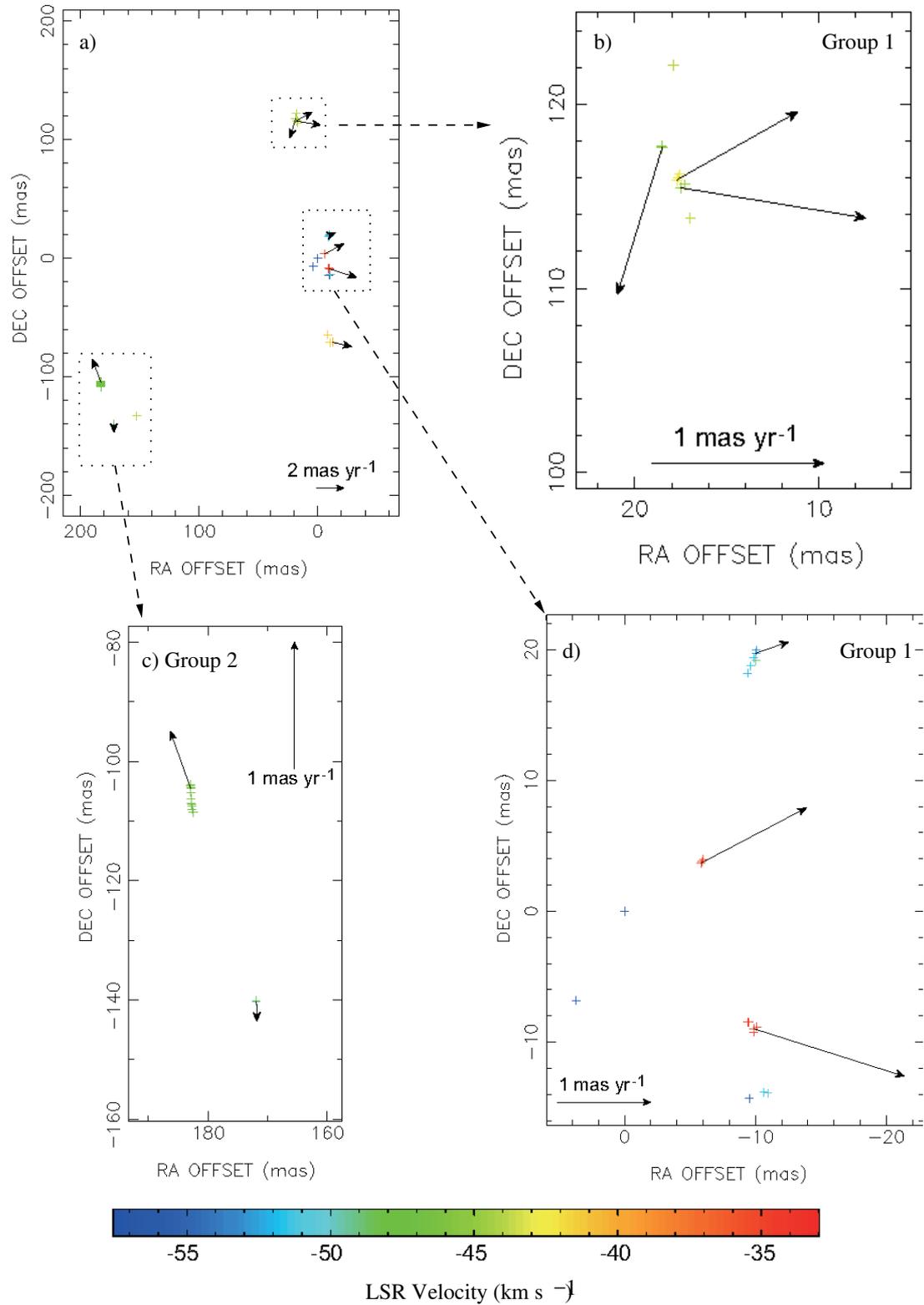}
 \end{center}
\caption{a) Proper motions of H$_2$O maser features detected towards Group 1 (dashed squares at the right) and Group 2 (dashed square at the left) for IRAS~23139+5939 (compare with figure \ref{fig:Epoch2}). Panel a) is referred to the dashed square inset shown in Fig. \ref{fig:Epoch2} and a close-up of groups 1 and 2 masers is shown in b), c), and d). {\bf A bias of 0.7~mas~yr$^{-1}$ toward north was subtracted from all maser proper motions.} For all panels, 1 mas yr$^{-1}$ = 16 km s$^{-1}$. }
 \label{fig:ProperMotions}
\end{figure*}

\begin{figure}
 \begin{center}
    \includegraphics[width=0.5\textwidth]{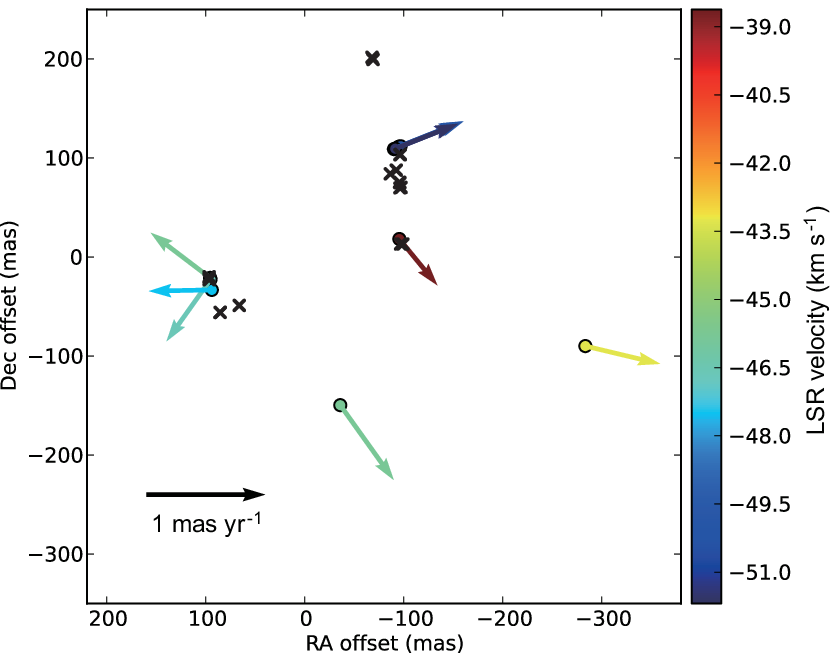}
 \end{center}
  \caption{Maser distribution found in this paper (cross symbols), which is overlaid with that (open circles) with the information of  maser proper motions (arrows; 1 mas yr$^{-1}$ = 16 km s$^{-1}$) found by \citet{Choi2014}. }
 \label{fig:Choi2014}
\end{figure}

Fig. \ref{fig:ProperMotions} shows proper motions, without the bias to the north, of H$_2$O maser features. We notice that the maser features in Group 1 are distributed within a region of $\sim$ 200 mas along the north-south direction around I23139 (Fig. \ref{fig:Epoch2-VLA}) and are moved essentially westwards.
In contrast to this, the maser features in Group 2 are aligned along the northeast--southwest direction and are moved northeast--southwestwards. This tendency reproduces clearly that scenario previously found by \citet{Choi2014} (see Fig. \ref{fig:Choi2014}). Moreover, the maser feature distribution seems to be consistent with its evolution between the epochs of the VLA \citep{Trinidad2006} and our KaVA observations (Fig. \ref{fig:Epoch2-VLA}). It exhibits east--west expansion between the maser features in Group 1 and 2 in the time baseline of $\sim$17~yr among these observations, which is consistent with that expected one from relative maser proper motions derived from this paper ($\sim$16~mas). However, given the positional uncertainty of KaVA observations, this result should be taken meticulously and must be confirmed with new observations.

\begin{longtable}{rr@{}rr@{}rr@{}r}
\caption{H$_2$O maser features identified in IRAS 23139+5939 with KaVA. }
\label{tab:tab2}
\hline
$V_{\rm LSR}$ & \multicolumn{2}{c}{R.A. offset$^{\ast}$} & \multicolumn{2}{c}{Decl. offset$^{\ast}$} & \multicolumn{2}{c}{$I_{\rm peak}$}  \\
\hspace*{\fill} (km s$^{-1}$) & \multicolumn{2}{c}{(mas)} & \multicolumn{2}{c}{(mas)} & \multicolumn{2}{c}{(Jy beam$^{-1}$)} \\
\endfirsthead
\hline
\endhead
  \hline
\multicolumn{7}{l}{$^{*}$Offset in J2000.} \\
\multicolumn{7}{l}{$^{\dagger}$Maser features used to determine the reference position.}  \\
\multicolumn{7}{l}{$^{\ddagger}$Maser feature including the velocity channel that was referenced in the fringe-fitting and self-calibration procedures.} \\
\multicolumn{7}{l}{$^{\star}$ Absolute coordinates: R.A.(J2000)$=$23$^{\rm h}$16$^{\rm m}$10$^{s}$\hspace{-2pt}.3684$\pm 0^{\rm s}$\hspace{-2pt}.0008, Dec (J2000)$=+$59$^{\circ}$55$^{\prime}$28$^{\prime\prime}$\hspace{-2pt}.492$\pm 0^{\prime\prime}$\hspace{-2pt}.060.}  \\
\endfoot
  \hline
  \multicolumn{7}{c}{2018 February 6} \\
  \hline
--44.04 & 17.88 & 0.09 & 122.13 & 0.05 & 0.582 & 0.053 \\
--45.73 & 18.49 & 0.10 & 117.67 & 0.08 & 0.447 & 0.051 \\
{$^{\dagger}$}--42.67 & 17.71 & 0.01 & 115.89 & 0.01 & 7.669 & 0.054 \\
--43.73 & 16.97 & 0.03 & 113.81 & 0.02 & 1.891 & 0.071 \\
--51.63 & --9.65 & 0.02 & 18.76 & 0.02 & 3.357 & 0.083 \\
{$^{\dagger}$}--35.20 & --5.84 & 0.03 & 3.67 & 0.02 & 1.759 & 0.070 \\
--34.35 & --9.38 & 0.08 & --8.48 & 0.06 & 0.567 & 0.054 \\
--32.98 & --9.88 & 0.12 & --9.24 & 0.10 & 0.346 & 0.055 \\
--51.63 & --10.63 & 0.06 & --13.84 & 0.06 & 1.062 & 0.089 \\
--41.20 & --8.37 & 0.15 & --64.69 & 0.04 & 0.885 & 0.077 \\
--40.04 & --12.56 & 0.03 & --70.92 & 0.02 & 2.384 & 0.082 \\
--41.41 & --10.35 & 0.01 & --71.09 & 0.01 & 6.782 & 0.082 \\
--46.57 & 182.89 & 0.02 & --104.58 & 0.02 & 3.709 & 0.082 \\
--46.47 & 182.79 & 0.03 & --106.29 & 0.05 & 1.744 & 0.082 \\
{$^{\ddagger}$}--46.89 & 182.74 & 0.01 & --107.45 & 0.01 & 9.539 & 0.082 \\
--47.31 & 182.66 & 0.03 & --108.08 & 0.03 & 2.183 & 0.080 \\
--47.52 & 182.49 & 0.06 & --108.54 & 0.06 & 1.099 & 0.080 \\
--48.99 & 171.89 & 0.03 & --140.28 & 0.03 & 1.381 & 0.058 \\
--41.62 & 801.39 & 0.10 & --460.94 & 0.05 & 0.675 & 0.060 \\
\hline
        \multicolumn{7}{c}{2018 March 29} \\
    \hline
--45.67 & 18.52 & 0.06 & 117.67 & 0.07 & 0.660 & 0.058 \\
{$^{\dagger}$}--42.62 & 17.65 & 0.01 & 116.04 & 0.01 & 4.574 & 0.058 \\
--44.51 & 17.46 & 0.03 & 115.45 & 0.04 & 1.304 & 0.058 \\
--53.04 & --10.02 & 0.01 & 19.71 & 0.01 & 5.189 & 0.075 \\
--51.89 & --9.86 & 0.01 & 19.35 & 0.01 & 4.892 & 0.072 \\
--49.15 & --10.03 & 0.05 & 19.14 & 0.07 & 0.696 & 0.052 \\
{$^{\dagger}$}--35.24 & --5.92 & 0.01 & 3.81 & 0.02 & 3.504 & 0.063 \\
{$^{\ddagger}$}--56.84 & --0.00 & 0.00 & --0.02 & 0.00 & 102.500 & 0.275 \\
--34.82 & --9.48 & 0.04 & --8.48 & 0.04 & 1.273 & 0.057 \\
--35.77 & --9.87 & 0.06 & --8.99 & 0.07 & 0.747 & 0.061 \\
--57.57 & --9.55 & 0.09 & --14.30 & 0.11 & 0.490 & 0.050 \\
--50.41 & --10.94 & 0.06 & --13.91 & 0.07 & 0.672 & 0.056 \\
--39.98 & --12.60 & 0.02 & --70.76 & 0.02 & 2.641 & 0.060 \\
--41.35 & --10.38 & 0.01 & --70.99 & 0.01 & 5.772 & 0.073 \\
--46.83 & 182.92 & 0.02 & --103.92 & 0.02 & 3.433 & 0.088 \\
--46.62 & 182.91 & 0.01 & --104.38 & 0.02 & 5.519 & 0.089 \\
--46.51 & 182.92 & 0.01 & --105.26 & 0.03 & 4.782 & 0.084 \\
--46.83 & 182.83 & 0.01 & --107.07 & 0.01 & 7.940 & 0.089 \\
--43.88 & 152.56 & 0.06 & --132.92 & 0.07 & 0.626 & 0.055 \\
--48.94 & 171.89 & 0.03 & --140.13 & 0.04 & 1.463 & 0.057 \\  
--41.67 & 801.32 & 0.02 & --460.84 & 0.03 &  1.708 & 0.062 \\  
--48.20 & 755.25 & 0.05 & --464.24 & 0.06 &  0.693 & 0.051 \\  
--46.94 & 754.49 & 0.07 & --464.68 & 0.09 &  0.746 & 0.072 \\
\hline
       \multicolumn{7}{c}{2018 April 22} \\
\hline
--45.58 & 18.54 & 0.05 & 117.73 & 0.06 & 1.984 & 0.213 \\
{$^{\dagger}$}--42.94 & 17.58 & 0.01 & 116.18 & 0.01 & 8.259 & 0.211 \\
--44.63 & 17.26 & 0.02 & 115.63 & 0.03 & 4.178 & 0.213 \\
--53.05 & --10.07 & 0.02 & 19.94 & 0.03 & 4.900 & 0.213 \\
--52.21 & --9.38 & 0.02 & 18.17 & 0.02 & 5.289 & 0.213 \\
{$^{\dagger}$}--35.36 & --6.00 & 0.06 & 3.97 & 0.08 & 1.819 & 0.220 \\
{$^{\ddagger}$}--56.64 & --0.00 & 0.00 & 0.01 & 0.00 & 66.630 & 0.358 \\
--56.42 & 3.70 & 0.02 & --6.83 & 0.04 & 4.984 & 0.358 \\
--34.73 & --10.10 & 0.04 & --8.85 & 0.05 & 2.703 & 0.222 \\
--39.99 & --12.72 & 0.03 & --70.71 & 0.03 & 3.587 & 0.203 \\
--41.36 & --10.39 & 0.02 & --70.98 & 0.02 & 6.740 & 0.216 \\
--46.73 & 182.96 & 0.03 & --104.15 & 0.08 & 2.837 & 0.260 \\
--46.84 & 182.77 & 0.02 & --107.25 & 0.03 & 7.173 & 0.297 \\
\hline
       \multicolumn{7}{c}{Reference position{$^{\star}$} }  \\
\hline
        & 5.87 & 0.01 & 59.93 & 0.02 &       &       \\
\hline
\end{longtable}

\section{Discussion}
\label{sec:disc}

In recent years, evidence that massive YSOs undergo an outflow stage governed by same processes as low-mass stars have increased. However, the detection of thermal jets has been very scarce. The aforementioned is particularly relevant since jets play a crucial role in forming low-mass stars, being the driving sources of molecular outflows (e.g. \cite{Anglada2018}). So, it is not still clear whether high-mass stars are formed in the same way as low-mass stars. In this way, it is necessary to carry out individual studies of massive YSOs. In particular, IRAS~23139$+$5939 is an excellent candidate to investigate whether the thermal jet could be driving the bipolar outflow observed in a large scale, where H$_2$O masers have been detected towards a possible thermal radio jet \citep{Trinidad2006}.
 
Making a comparison with other massive star-forming regions (e.g., Cepheus~A, AFGL~2591; \cite{Torrelles2001, Torrelles2011, Trinidad2013}), there is a relatively small number of H$_2$O maser features detected toward IRAS~23139$+$5939 (see Figure \ref{fig:Epoch2}), and most of them do not persist over timescales of a few weeks. As a result, it has been challenging to find a regular and systematic maser distribution and measure a significant number of proper motions. Because of this reason, a qualitative description is only given. However, we newly found that the groups of H$_2$O maser features have persisted since the previous VLA observations (see Fig. \ref{fig:Epoch2-VLA}). It allows us to track the evolution of the maser distribution and discuss further the properties of the H$_2$O masers concerning I23139 and the morphology of the outflow from this YSO in more detail. 

The relative positions of the H$_2$O maser features concerning the 1.3~cm continuum source I23139 have an accuracy of 10 mas in the VLA data \citep{Trinidad2006}. Considering that the maser features in Group 1 observed with KaVA are also spatially associated with those found in the VLA observation, these KaVA maser features are also associated with I23139, being this the pumping source. \citet{Trinidad2006} found that the VLA maser features spatially associated with I23139 are not gravitationally bound with this source but rather tracing expansion motions. Measured maser proper motions prove this suggested expansion. 

\begin{figure}
 \begin{center}
  \includegraphics[width=0.5\textwidth]{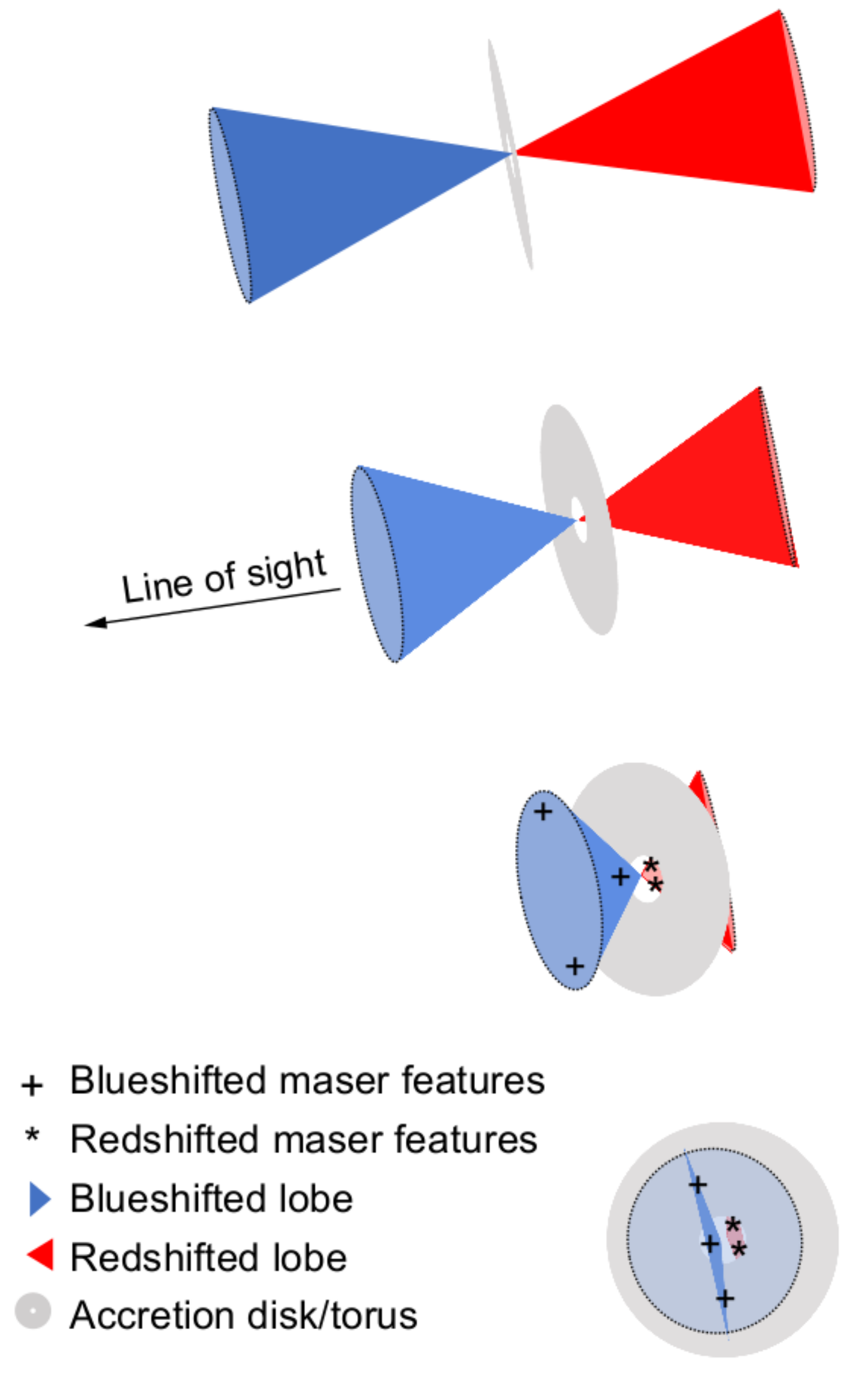}
  \end{center}
 \caption{Cartoon of the disk-YSO-outflow system in I23139 in four viewing angles. Observed masers may suggest the third or fourth case of the viewing angle. Under this scenario,  blue-shifted maser components appear widely in the region, while most of  red-shifted components behind  disk/torus are invisible, except those located closely to the central driving source.}
 \label{fig:cartoon}
\end{figure}

From Table \ref{tab:tab2}, we notice that the strongest H$_2$O maser feature is found in the central subgroup of Group 1, spatially coinciding with the continuum source I23139. In this subgroup,  maser features have the lowest and highest radial velocities, i.e. those with the highest blue- and red-shifts, respectively. In particular, for the second session of  KaVA observations,  maser features with ``blue'' velocities of the central subgroup were more intense (up to tens of Jy) than those with ``red'' velocities (a few tenths of Jy). The latter could indicate that the ``blue'' maser features are in front of the continuum source, while the ``red'' ones are behind it, so we speculate these look like less intense due to ``red'' maser emission is being masked by a dense structure, for example, a circumstellar disk (see Fig. \ref{fig:cartoon}).

In the same way, for the maser source G353.273$+$0.641, \citet{Motogi2011} proposed a disk-masking scenario to explain the nature of the strong blueshift dominance, where an optically-thick disk obscures a redshifted lobe of a compact jet. Under this geometry, we expect an accretion disk/torus around I23139 in the plane of the sky, perpendicular to the line-of-sight (pole-on). Herein, the H$_2$O maser emission, which is also projected onto the plane of the sky, could be tracing a jet (or, more generally, a conical outflow). In this case,  we can also expect the velocities of the H$_2$O maser features to increase when these are moved away from I23139.

Considering that the spatial distribution of H$_2$O maser features in Group 1 is extended over a scale of $\sim$200 mas, we expect an optically thick disk of about 300~AU. 
Fig. \ref{fig:cartoon} shows the schematic summary of our consideration in terms of the outflow morphology hosting the H$_2$O masers in I23139. This scenario seems to be consistent with the spatial distribution, velocity field, and intensities of the H$_2$O maser features in Group 1 and their expanding motions. However, given the angle of inclination of the outflow in the proposed scenario and few calculated proper motions, we cannot perform a detailed study. 

Despite these limitations, this scenario also seems to be congruent with the observed spatial and kinematic distribution of a wide-angle molecular outflow toward IRAS~23139$+$5939. This outflow is closely oriented to the line-of-sight with extended high-velocity wings  \citep{Wouterloot1989,Beuther2002,Sanchez2013}, where emission from blueshifted wings is more intense than those redshifted wings.

Besides, \citet{Trinidad2006} suggested that I23139 is a thermal radio jet associated with a massive YSO, while the H$_2$O maser distribution, which we found, predicts the presence of an accretion disk around I23139. These results show that I23139 is forming a disk-YSO-jet system similar to that observed in low mass YSOs. Under these conditions, I23139 could be an example of a high-mass protostar with a disk-YSO-jet system, suggesting that I23159 is being formed by accretion.

As it was previously indicated, the spatial extension of maser features is about 200 mas. Assuming that these features are tracing a wide-angle outflow (poorly collimated) and expanding with a velocity of $\sim$ 45 km s$^{-1}$ (obtained using their radial velocity and proper motions), we found a rough estimation of the dynamical age of 35 yr for the ejection traced by maser features. This dynamical age is greater than the calculated for W75-VLA 2 \citep[13 yr]{Torrelles2003} but comparable with Cepheus A HW2-R5 \citep[33 yr]{Torrelles2001}.

In both cases, these values have been interpreted as short-lived episodic spherical winds from a massive object located at the center of the structure, which occurs during the early stages of their evolution. Although the dynamical age obtained for the outflow traced by the maser features is consistent with those obtained in other young protostars, this value must be taken  carefully because the H$_2$O maser outflow seems to be closely oriented to line-of-sight, and its spatial extension could be uncertain. Either way, if the estimated age is correct, this result could suggest that the continuum source, I23139 (spatially associated with the maser features of group 1), is in an early evolutionary stage of its formation, which is consistent with it by being a radio jet.

In addition, we notice that the flux density of the continuum source I23139 also shows variability in time.  With VLA observations in the 22~GHz band (both in the A configuration), it had a flux density of 0.98 mJy in 2001 \citep{Trinidad2006} but 0.25 mJy 12 years later \citep{Moscadelli16}. A similar behavior is observed in the 8.4~GHz band (0.92 and 0.53 mJy; see \cite{Tofani1995} and \cite{Trinidad2006}, respectively). However, the spectral index estimated for I23139 between 8.4 and 22.2~GHz, with simultaneous observations \citep{Trinidad2006}, is very similar to that estimated between 15 and 22~GHz \citep{Moscadelli16}, 0.64$\pm$0.36 and 0.72$\pm$0.30, respectively. The difference in the measured flux density could be attributable to the intrinsic variability of the continuum source, which could contribute to the variability of the H$_2$O masers detected in the region. Furthermore, we notice that the flux density of the H$_2$O maser emission is correlated with the intensity of the continuum source at the 22~GHz band. This possible correlation may be further tested by revisiting this maser source in the future. 

Regarding the maser features of Group 2, \citet{Trinidad2006} found that these masers trace a large filament along the northeast-southwest direction and seem to be pointing toward the 3.6~cm continuum source I23139S, located 0.35\arcsec\ to the southwest from these masers and interpreted as a thermal jet. \citet{Trinidad2006} speculated that the H$_2$O maser features could be associated with and excited by this continuum source.
Although in general, the distribution of H$_2$O masers observed with the VLA is maintained in KaVA observations, only those that are further north were detected. However, the directions of the H$_2$O maser proper motions do not seem to be consistent with I23139S being the driving source. Based on our observations, we found that  maser features are roughly aligned with I23139S, but these point in the opposite direction, which rules out that I23139S is the exciting source.
Besides, these maser features seem to be tracing an expansion motion driven by an undetected young source. This scenario has also been found in other star-forming regions, for example in AFGL~2591~VLA˜3--N \citep{Trinidad2013}, although in this case, the maser features trace an expanding spherical movement rather than a collimated motion.

\citet{Moscadelli16} found a second weak continuum source in the 22~GHz band, which is close to the midpoint of the Group 2 of maser features detected with the VLA and KaVA. This continuum source could be a candidate for the driving source for this group of H$_2$O masers. However, this new continuum source has only been detected at 22~GHz, so it is not still possible to calculate its spectral index to research its nature. 
Nevertheless, using the flux density of the continuum source at 22~GHz and an upper limit for its flux density at 8.4~GHz (3$\sigma$; \cite{Trinidad2006}), we estimate a rough spectral index $\lesssim -$0.17, which could be consistent with an optically thin \hii region. If this continuum source is responsible for exciting H$_2$O masers, these should trace a spherical expansion and not a collimated motion. Despite the proper motions of the maser features in Group 2 point in the opposite direction, their velocities are blueshifted, indicating that these features are in front of the continuum source and
are moved on the plane of the sky. Under this scenario, the H$_2$O maser features may be on the surface of the \hii region and pumped by the central star (see Fig. \ref{fig:Epoch2-VLA}).

\begin{table*}
\centering
 \begin{minipage}{140mm}
    \caption{Parameters of the H$_2$O maser features with proper motions measured with KaVA in IRAS~23139+5939 }
      \label{tab:tab3}
\begin{tabular}{@{}ccccrrrr@{}}
 \hline
Feature & Detected & \multicolumn{1}{c}{$V_{\rm LSR}$} & 
\multicolumn{1}{c}{$I_{\rm peak}$} & \multicolumn{1}{c}{$\Delta x$} & 
\multicolumn{1}{c}{$\Delta y$} & \multicolumn{1}{c}{$V_{\rm x}$} & 
\multicolumn{1}{c}{$V_{\rm y}$}  \\
ID & epochs & (km s$^{-1}$) & (Jy beam$^{-1}$)) & \multicolumn{1}{c}{(mas)} & 
\multicolumn{1}{c}{(mas)} & \multicolumn{1}{c}{(km s$^{-1}$)} & 
\multicolumn{1}{c}{(km s$^{-1}$)}  \\
    \hline
1 & 1,2,3 & --45.67 & 1.98 &  18.52$\pm$0.06 & 117.67$\pm$0.07 &  2.74$\pm$6.26 & --13.16$\pm$2.47 \\
2 & 1,2,3 & --42.62 & 8.26 & 17.65$\pm$0.01 & 116.04$\pm$0.01 & --7.34$\pm$0.79 & --0.07$\pm$3.77 \\
3 & 2,3 & --44.51 & 4.18 & 17.46$\pm$0.03 & 115.45$\pm$0.04 & --21.79$\pm$3.93 & 3.51$\pm$1.06 \\
4 & 2,3 & --53.04 & 5.19 & --10.02$\pm$0.01 & 19.71$\pm$0.01 & --5.45$\pm$2.44 & 8.96$\pm$0.94 \\
5 & 1,2,3 & --35.24 & 3.50 & --5.92$\pm$0.01 & 3.81$\pm$0.02 & --9.02$\pm$3.76 & 0.52$\pm$3.82 \\
6 & 2,3 & --35.77 & 2.70 & --9.87$\pm$0.06 & --8.99$\pm$0.07 & --25.06$\pm$7.86 & --0.85$\pm$4.99 \\
7 & 1,2,3 & --39.98 & 3.59 & --12.60$\pm$0.02 & --70.76$\pm$0.02 & --9.02$\pm$2.38 & --6.28$\pm$0.20 \\
8 & 1,2,3 & --46.62 & 5.52 & 182.91$\pm$0.01 & --104.38$\pm$0.02 & 3.86$\pm$2.02 & 6.55$\pm$3.37 \\
9 & 1,2 & --48.94 & 1.46 & 171.89$\pm$0.03 & --140.13$\pm$0.04 & --0.12$\pm$4.89 & 0.88$\pm$1.37 \\    
   \hline
\end{tabular}
\noindent {\bf Notes}: Column 1 gives the identification number of the H$_2$O maser feature. Column 2 gives the different epochs in which the maser feature was detected (epoch 1: 2018 February 6; epoch 2: 2018 March 29; 2018 3: 2002 May 22). Column 3 gives the mean local-standard-of-rest (LSR) velocity of the maser feature. Column 4 gives the highest peak intensity of the maser feature in the three epochs. Columns 5 and 6 give the position offset of the feature (referred to as the second epoch). Columns 7 and 8 give the components of the transverse velocity, which is converted from the proper motion using a source distance of 3.4~kpc. The position of the ``center of motion'' is ($2.17\pm0.03$, $41.85\pm0.03$) (($V_{\rm x}$, $V_{\rm y}$) = (0, 0)~km s$^{-1}$). 
\end{minipage}
\end{table*}

\section{Conclusions}
\label{sec:summary}

Using the data of multi-epoch observations of H$_2$O masers with KaVA, we detected three groups of masers towards IRAS 23139+5939.
Only Groups 1 and 2 coincide with the continuum source I23139. KaVA observations newly found Group 3, which was not detected in the previous VLA observations and it is located quite far from I23139 (approximately 0.9\arcsec). The H$_2$O maser emission toward IRAS  23139+5939  is highly variable with time.  Because of this reason, very few maser features persist on time scales of several weeks,  making it difficult to determine proper motions in a single set of VLBI observations.

Based on the proper motions of maser features, we can confirm that  H$_2$O masers, which are associated with the continuum source I23139 (Group 1), are tracing expansion motions, possibly a wide-angle short-lived outflow with a rough dynamical age of 35 yr. Furthermore, using a very simplistic model, we suggest that these masers should be tracing an outflow seen from the front, where only masers located at the outflow's base and on the outermost part of it are detected. Moreover, this model predicts the presence of a circumstellar disk around I23159, suggesting that this massive protostar is in an early evolutionary stage and forming in a similar way to low-mass stars. Finally, we found that maser features in the Group 2 seem to be tracing expansion motions on the surface of an \hii region.


\begin{ack}

VERA is operated by Mizusawa VLBI Observatory, respectively, branches of the National Astronomical Observatory of Japan, National Institutes of Natural Sciences. KVN in KaVA and the Daejeon Correlator in KJCC is operated by Korea Astronomy and Space Science Institute (KASI). MAT acknowledges support from DAIP-UG grant 163/2021. HI was supported by the JSPS KAKENHI Grant Number JP16H02167 and JP21H00047. IT--J acknowledges support from CONACyT, Mexico; grant 754851. EdelaF thanks Kagoshima University for financial support during the visit to the CGE, Institute for Comprehensive Education, Kagoshima University in 2019. He also renders thanks to UdG-PRO-SNI 2021 grant, the staff of the Institute for Cosmic Ray Research (ICRR-UTokyo), Coordinaci\'on General Acad\'emica y de Innovación (CGAI-UDG), Coordinaci\'on de Servicios Acad\'emicos (CSA-CUCEI-UDG), PRODEP-SEP through Cuerpo acad\'emico UDG-CA-499, Carlos Iv\'an Moreno, Cynthia Ruano, Rosario Cedano, Diana Naylleli Navarro, Luis Fernando Gonz\'alez Bola\~nos, Jorge Alberto Rodr\'iguez Castro, Johnatan Barcenas Alvarez, Patricia Retamoza Vega, Laura Gonzalez Jaime, and Dulce Ang\'elica Valdivia Ch\'avez, Kumiko Sugimoto, and Michiru Ito for financial and administrative supports during Sabbatical year stay at the ICRR, the University of Tokyo in 2021. 

\end{ack}

\end{document}